\documentclass[sigconf]{acmart}
\AtBeginDocument{%
  }

\acmSubmissionID{7673}

\usepackage{pifont}
\newcommand{\cmark}{\ding{51}}% ✓
\newcommand{\xmark}{\ding{55}}% ✗

\begin{document}

\title{GlobalizeEd: A Multimodal Translation System that Preserves Speaker Identity in Academic Lectures}

%%%%%%%%%%%%%%%% AUTHORS %%%%%%%%%%%%%%%%
%%%%%%%%%%%%%%%%%%%%%%%%%%%%%%%%%%%%%%%%%%

\author{Hoang-Son Vo$^*$}
\affiliation{
  \institution{Chonnam National University}
  \country{Republic of Korea}
}
\email{hoangsonvothanh@jnu.ac.kr}

\author{Karina Kolmogortseva$^*$}
\affiliation{
  \institution{Chonnam National University}
  \country{Republic of Korea}
}
\email{karinenish99@jnu.ac.kr}

\author{Ngumimi Karen Iyortsuun}
\affiliation{
  \institution{Chonnam National University}
  \country{Republic of Korea}
}
\email{karen@jnu.ac.kr}

\author{Hong-Duyen Vo}
\affiliation{
    \institution{FPT University}
    \country{Viet Nam}
}
\email{duyenvth4@fe.edu.vn}

\author{Soo-Hyung Kim$^\dagger$}
\affiliation{
    \institution{Chonnam National University}
    \country{Republic of Korea}
}
\email{shkim@jnu.ac.kr}

\thanks{$^*$ These authors contributed equally to this work.}
\thanks{$^\dagger$ Corresponding author.}


%%%%%%%%%%%%%%%% ABSTRACT %%%%%%%%%%%%%%%%
%%%%%%%%%%%%%%%%%%%%%%%%%%%%%%%%%%%%%%%%%%
\begin{abstract}
    A large amount of valuable academic content is only available in its original language, creating a significant access barrier for the global student community. This is a challenge for translating in several subjects, such as history, culture, and the arts, where current automated subtitle tools fail to convey the appropriate pedagogical tone and specialized meaning. In addition, reading traditional automated subtitles increases cognitive load and leads to a disconnected learning experience. Through a mixed-methods study involving 36 participants (18 instructors and 18 students), we found that GlobalizeEd's dubbed formats significantly reduce cognitive load and offer a more immersive learning experience compared to traditional subtitles. Although learning effectiveness was comparable between high-quality subtitles and dubbed formats, both groups valued GlobalizeEd’s ability to preserve the speaker’s voice, which enhanced perceived authenticity. Instructors rated translation accuracy (M = 4.22) and vocal naturalness (M = 3.67), whereas students reported that synchronized, identity-preserving outputs fostered engagement and trust. This work contributes a novel human-centered AI framework for cross-lingual education, demonstrating how multimodal translation systems can balance linguistic fidelity, cultural adaptability, and user control to create more inclusive global learning experiences.

\end{abstract}



%%%%%%%%%%%%%%%%%%KEY WORDS%%%%%%%%%%%%%%%%%%
\begin{teaserfigure}
  \includegraphics[width=\textwidth]{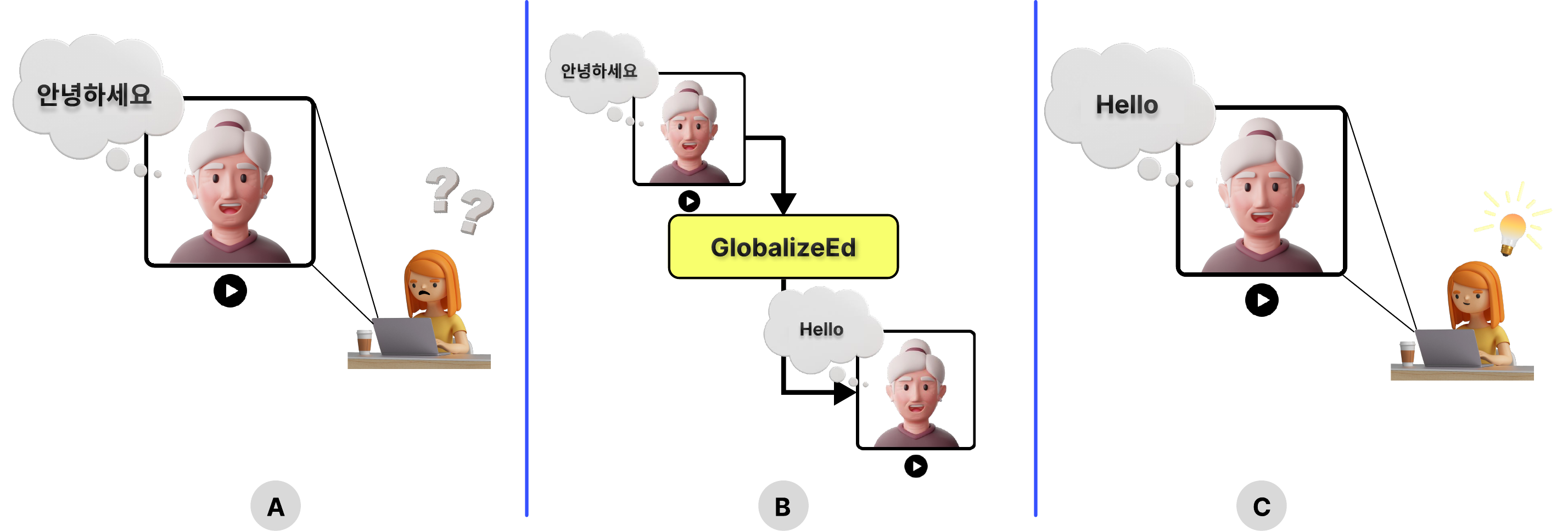}
  \caption{An illustration of the GlobalizeEd system. (A) A student struggles to watch a video in a different language. (B) Our system processes the video through a multimodal pipeline that translates the content, clones the voice, and synchronizes the lip movements. (C) The student can now understand the lecture through an authentic and accessible dubbed version.}
  \label{fig:teaser}
\end{teaserfigure}



%%%%%%%%%%%%%%%% CONTENT %%%%%%%%%%%%%%%%
%%%%%%%%%%%%%%%%%%%%%%%%%%%%%%%%%%%%%%%%%%
\maketitle

\section{Introduction}
\label{sec:intro}

The globalization of digital media has increased the demand for video translation systems that provide not only linguistic equivalence but also cultural and temporal coherence.  Platforms such as YouTube and TikTok are showing increasing interest in cross-cultural video consumption, but existing translation technologies still struggle to deliver outputs that are both culturally appropriate and temporally coherent. Traditional systems, which often use automatic subtitle generation or speech-to-speech pipelines, often suffer from three disadvantages: (i) temporary inconsistency when the translated speech exceeds or falls short of the length of the original video, resulting in poor audiovisual synchronization; (ii) lack of cultural adaptation when direct reproduction of the translated speech does not correspond to reality. The translations do not reflect differences in tone, politeness levels, or style of presentation in different languages; and (iii) limited user-centered control, which gives viewers or content creators little opportunity to adapt the results to contextual needs.

Latest advances in dubbing technologies have improved technical aspects such as lip synchronization and duration control, but most systems focus on model-centric optimization while neglecting the user-facing dimension of dubbing systems. Meanwhile, current studies such as Translation in the Wild~\cite{balashov2025translation} have shown that large language models (LLMs) possess strong translation abilities even without task-specific training, due to their exposure to large-scale multilingual and multimodal data. This makes them particularly suitable for flexible, context-aware translation tasks that require preserving tone, style, and cultural intent.

However, technical translation capability alone does not guarantee successful 
cross-cultural communication. Yet, studies in cross-cultural communication show that the success of media translation hinges not only on linguistic fidelity but also on cultural adaptability and audience perception. For instance, research on Chinese variety shows on YouTube found that international reception depends on genre innovation, tone adaptation, and the ability to transform "cultural discount" into "cultural premium"~\cite{zhang2025power}. These insights underline the necessity of systems that empower end-users to control translation style and tone, ensuring both intelligibility and cultural resonance.

In this paper, we present a multimodal video translation prototype designed to address these limitations through three innovations:
\begin{itemize}
    \item \textbf{Tone adaptation} – enabling users to select stylistic modes (e.g., formal, informal, friendly) to match cultural and situational context.
    \item \textbf{Voice cloning and lip synchronization} – preserving speaker identity and naturalness in translated outputs.
    \item \textbf{Duration alignment} – a lightweight alignment method that stretches or compresses translated speech to match the temporal structure of the source video.
\end{itemize}

We conducted a user study with English teachers and their students to assess the system. The participants translated educational videos into English using our system and assessed the resulting outputs in terms of naturalness, synchronization quality, comprehension, and cultural fit. Our findings demonstrate that duration alignment significantly improves perceived synchronization, while tone adaptation enhances cultural appropriateness, especially in educational contexts.

\subsection*{Contributions}
This work makes three main contributions:
\begin{itemize}
    \item We propose a culturally adaptive video dubbing system that integrates multimodal translation, style control, voice cloning, lip synchronization, and duration alignment.
    \item We provide one of the first user-centered evaluations of such a system, with quantitative and qualitative evidence from teachers and students.
    \item We situate our system within the broader trajectory of research on multimodal translation, dubbing, and cross-cultural communication, extending prior model-focused work into a practical, user-facing prototype.
\end{itemize}

By focusing on cultural adaptability and temporal alignment as first-class objectives, our work contributes to the emerging field of intelligent user interfaces for cross-cultural media consumption.

\section{Related Work}
\label{sec:relatedworks}

Lately research on video dubbing and multimodal machine translation (MMT) has sped up. This is because both academics and businesses want videos to be accessible across cultures.  We review prior  work relevant to our system, covering multimodal translation, duration alignment methods, pipeline integration, and user-centered design considerations.

\subsection{Multimodal Machine Translation}
Multimodality has emerged as a fundamental aspect of contemporary translation research.  Reviews and bibliometric studies highlight the shift from text-only pipelines to systems that integrate visual and auditory cues to improve alignment, disambiguation, and naturalness~\cite{shen2024survey,guo2025mapping}. These works emphasize that the use of multiple modalities not only enhances semantic accuracy but also makes user experience in audiovisual media better.

An example is the MAVL dataset, which provides multilingual audio-video lyrics matching with the musical rhythm~\cite{cho2025mavl}. This work demonstrates that good translation is not enough if it ignores timing and prosody,  when multimodal constraints shape user's perception. Similarly, efforts in unsupervised multimodal translation have used images as pivots for low-resource language pairs~\cite{tayir2024visual}, illustrating the role of multimodality in extending translation to underrepresented languages. Together, these studies establish multimodality as essential for moving beyond purely textual equivalence toward natural audiovisual adaptation.

\subsection{Duration Alignment in Video Dubbing}
Aligning the translated speech with the original video's temporal structure is a major issue in video dubbing. Duration mismatches are frequently caused by different information densities in different languages; translations may be noticeably longer or shorter than the source speech, which throws off the audio-visual coherence. This issue has been the focus of modern studies at two points in the dubbing process: text-to-text translation and text-to-speech speech synthesis.

\textbf{Translation-Stage Approaches.}
Several methods have been developed to create translations taking into account the duration at the stage of machine translation. Microsoft's Length-Sensitive Speech Translation (LSST) introduced phoneme-based length control in end-to-end speech translation~\cite{chadha2025length}. The model generates multiple translation candidates (short, normal, long) using predefined length tokens, then selects the variant closest to the source audio duration through a TTS duration estimator. Their Length-Aware Beam Search (LABS) algorithm efficiently produces all three variants in a single decoding pass, achieving 16.3\% improvement in Speech Rate Compliance for Spanish and 19.9\% for Korean while maintaining comparable BLEU performance. However, LSST operates on speech-to-text translation without visual modality, limiting its applicability to scenarios requiring lip synchronization.

Duration alignment was treated as a preference optimization problem by Segment Supervised Preference Optimization (SSPO) in this study~\cite{cui2025fine}. Unlike conventional methods that control overall text length, SSPO performs fine-grained, line-by-line duration matching during the translation stage. The method samples multiple translation candidates per dialogue line and selects preferred/rejected pairs based on a duration consistency metric measured via TTS synthesis. SSPO maintained translation quality across Chinese-English and Chinese-Thai pairs while achieving major improvements in duration consistency (P metric decreased by 30–40\%) through the use of Direct Preference Optimization (DPO) with segment-wise sampling. Critically, SSPO only requires ~3\% of training data (600 prompt-response pairs) to achieve notable improvements. Nevertheless, both LSST and SSPO focus exclusively on duration metrics without addressing emotional tone, cultural adaptation, or user-facing style control.

\textbf{Synthesis-Stage Approaches.}
At the speech synthesis stage, recent work has explored video-guided and multimodal techniques for duration control. FlowDubber proposed combining LLM-based semantic-aware learning with flow matching for high-quality dubbed speech generation~\cite{cong2025flowdubber}. The system uses Qwen2.5 as a speech language model to capture phoneme-level pronunciation while employing Dual Contrastive Alignment (DCA) to ensure fine-grained synchronization between phoneme sequences and lip movements extracted from video. Their Flow-based Voice Enhancing (FVE) mechanism improves acoustic quality through LLM-guided flow matching and style-aware transformations. FlowDubber achieved state-of-the-art lip synchronization (LSE-C/LSE-D metrics) on Chemistry and GRID benchmarks while maintaining high UTMOS scores for acoustic quality. However, the approach requires substantial computational resources for flow-based generation and operates after translation is complete, lacking integrated control over translation length or style.

DubWise, developed by Sony Research, introduced explicit video-guided duration control in autoregressive TTS~\cite{sahipjohn2024dubwise}. The system integrates lip-reading features extracted via AV-HuBERT into a GPT2-based TTS model (XTTS backbone) using cross-modal attention layers. By conditioning on cropped lip regions from reference video, DubWise generates speech aligned with observed lip movements even when the input text differs from the original or is in a different language. The method includes a novel duration loss term that explicitly penalizes mismatches between predicted and ground-truth sequence lengths. DubWise demonstrated improved duration ratio (DR closer to 1.0) and lower Word Error Rate compared to baselines like FastSpeech2 and YourTTS across same-text, different-text, and cross-lingual scenarios on Lip2Wav-Chemistry and LRS2 datasets. Notably, the approach trains only cross-attention and transposed convolution layers, making it computationally lightweight. In contrast to FlowDubber, DubWise assumes pre-translated text as input and lacks mechanisms for cultural or stylistic adaptation beyond voice cloning.

Collectively, these works demonstrate that duration alignment is critical for perceived dubbing quality, but they address different pipeline stages and optimization objectives. Translation-stage methods (LSST, SSPO) control text length before synthesis, while synthesis-stage methods (FlowDubber, DubWise) adjust speech generation to match video constraints.

\begin{table*}[t]
\centering
\caption{Comparison of Approaches to Video Dubbing and Translation}
\label{tab:comparison}
\scriptsize
\setlength{\tabcolsep}{3pt}
\renewcommand{\arraystretch}{1.2}

\begin{tabular}{|p{2.6cm}|p{2.6cm}|p{2.2cm}|p{1.3cm}|p{2.1cm}|p{3.4cm}|}
\hline
\textbf{System} & \textbf{Pipeline Stage} & \textbf{Duration Control} & \textbf{Lip Sync} & \textbf{User Control} & \textbf{Focus / Contribution} \\
\hline
FlowDubber & TTS (speech synthesis) & \cmark{} (duration predictor + DCA) & \cmark{} & \xmark{} & High A/V sync + acoustic quality via flow matching \\
\hline
MAVL & Benchmark (lyrics MT) & \cmark{} (syllable-constrained) & \xmark{} & \xmark{} & Multimodal dataset for singable lyrics translation \\
\hline
LSST & ST (speech$\rightarrow$text) & \cmark{} (3 length variants) & \xmark{} & \xmark{} & Real-time/on-device ST with phoneme-based control \\
\hline
SSPO & MT (text$\rightarrow$text) & \cmark{} (segment-level DPO) & \xmark{} & \xmark{} & Preference optimization for line-level duration \\
\hline
DubWise & TTS (multimodal) & \cmark{} (video-guided) & \cmark{} & \xmark{} & Video-conditioned TTS with cross-modal attention \\
\hline
\textbf{Our System} & \textbf{Full pipeline} & \cmark{} (adaptive) & \cmark{} (optional) & \cmark{} (tone/style UI) & \textbf{End-to-end deployment with user control}\\
\hline
\end{tabular}
\end{table*}

\subsection{Pipeline Integration and Practical Systems}

While most research focuses on optimizing individual pipeline stages (ASR, MT, or TTS), fewer works address end-to-end integration for practical deployment. Traditional cascaded systems suffer from error propagation and lack of global optimization across stages.

Recent industrial efforts have explored tighter integration. For example, SeamlessM4T provides speech-to-speech translation with prosody preservation but does not address video alignment~\cite{barrault2023seamlessm4t}. VideoDubber approached the problem from the MT perspective, proposing translation systems conditioned on reference speech duration~\cite{wu2023videodubber}.

Our work bridges this gap by providing a user-facing system that integrates: (1) automatic subtitle generation, (2) duration-aware translation, (3) style-controlled voice cloning, and (4) optional lip synchronization. Unlike research prototypes optimizing specific metrics, we prioritize usability, allowing non-expert users to control tone and style through simple UI selections.

\subsection{User Experience and Cross-Cultural Communication}
While much research emphasizes technical synchronization, fewer works explore the role of cultural adaptation in dubbing systems. Yet cross-cultural communication studies demonstrate that reception depends not only on fidelity but also on adaptability.

For example, an analysis of Chinese variety shows on YouTube found that audience engagement and acceptance were strongly influenced by genre innovation, tone adaptation, and the ability to transform cultural "discounts" into cultural "premiums" ~\cite{zhang2025power}. These findings align with our focus on tone adaptation as a core feature of user-facing video translation. By allowing users to select stylistic modes—formal, informal, or friendly—our system enables translations to resonate with different cultural and situational contexts.

\subsection{Toward Multimodal Interaction Systems}
Beyond translation and dubbing, the broader multimodal interaction community has begun exploring real-time, memory-augmented systems for continuous video and audio processing. For instance, InternLM-XComposer2.5-OmniLive introduces a streaming perception, memory, and reasoning architecture for long-term multimodal interaction~\cite{zhang2024internlm}. Such systems highlight future directions for translation technologies: real-time streaming, context-aware memory, and adaptive reasoning.

At the same time, ethical considerations remain critical. Recent research has shown that automated video-based LLM annotation can raise privacy concerns, and watermarking-based protections have been proposed to disrupt misuse~\cite{liu2025protecting}. These insights underscore the importance of embedding privacy and ethical safeguards in user-facing dubbing systems, particularly as they scale to global media platforms.

\section{System Design}
\label{systemdesign}

To solve the problems that we discussed, we built a system called GlobalizeEd. It is a tool that can translate and dub academic videos. When we designed the system, we had three main goals. First, we wanted the translation to be accurate and to understand the topic well. This is important to keep the original meaning of the lecture. Second, we wanted to keep the instructor's original voice and accent. This makes the video feel more authentic, not synthetic or robotic. Third, we wanted the system to be accessible to instructors with varying levels of technical proficiency.

Our system uses several AI technologies together in a new way to work for education. It is different from other systems because it focuses on the experience of the user, such as reducing the difficulty of learning and keeping the instructor's teaching style. In this section, we will explain how our system works. First, we will show the overall process from start to finish. Then, we will talk about the main AI parts inside the system. Finally, we will show the user interface that we designed for teachers.

\subsection{System Architecture}

\begin{figure*}[h]
  \centering
  \includegraphics[width=\linewidth]{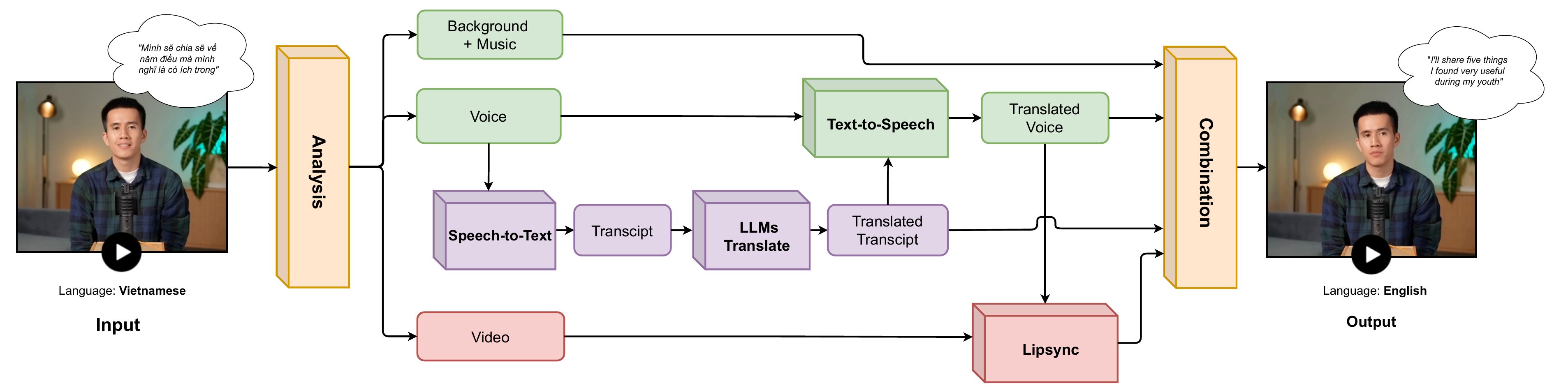}
  \caption{The AI system architecture of the GlobalizeEd multimodal translation pipeline. The input video is analyzed to separate its components: the speaker's voice, background audio, and the video stream. The voice is transcribed via Speech-to-Text, and an LLM translates the resulting transcript. The translated transcript is then put through the Text-to-Speech, which mimics the original voice accent to produce a new audio. This new audio and the original video stream are then fed into a lip-sync module to generate synchronized mouth movements. Finally, all components of the background audio, the translated voice, and the lip-synced video are combined to produce the final dubbed video output.}
  \label{fig:system}
\end{figure*}

Our system was developed to transform a video lecture from a source language into a fully translated and synchronized version in a target language. Figure \ref{fig:system} shows a high-level overview of this architecture. The entire process is designed to be a logical flow, breaking down a complex task into a series of smaller, manageable steps. We will explain each major stage of this pipeline below.

\subsubsection{Analysis and regeneration of components}
\label{sec:sub:analysis}

When an instructor uploads a video, the first step is Analysis. This stage acts like a sorting mechanism for the video's different data streams. A typical academic video contains more than just the speaker's voice; it often includes background music or sound effects. To handle these properly, our system separates the input video into three main components:

\begin{itemize}
    \item The Video Stream: This is the visual part of the lecture, containing the footage of the instructor. This stream is kept separate so we can later modify the instructor's lip movements without affecting the rest of the video.

    \item The Primary Voice Stream: This is the most important audio component-the instructor's spoken words. We use an audio separation tool to isolate the vocal track from other sounds. This clean vocal track is crucial for the next steps.

    \item The Background Audio Stream: This includes everything else, such as background music, presentation sounds, or ambient noise. We keep this track to add it back at the end, which helps the final video feel natural and not empty.
\end{itemize}

Once the voice track is isolated, it is processed to generate a transcript. The clean voice audio is sent directly to a Speech-to-Text module, which creates a written transcript of the lecture. This transcript contains the original words and their timestamps.

Therefore, after the processing, we produce 4 main tracks, including the video stream, voice track of the lecture, background audio, and the transcript.

\subsubsection{LLM-Powered Translation}

The core of our system is the translation module. For the final translated video to feel natural, the translation must be more than just accurate. It needs to capture the original context and tone to be suitable for spoken delivery, allowing it to adapt to the cultural nuances of the target language. This means the text must be conversational in style and align with the timing of the original lecture.
    
To achieve this, we applied Prompt Engineering to guide a Large Language Model (LLM). Instead of simply asking the model to translate the entire transcript at once, we process the text segment by segment. We give the model a detailed set of instructions for each segment, which frames the task in the specific context of video content. Our prompt uses two main strategies to get a better result.

First, we use Context Augmentation \cite{kobayashi-2018-contextual} to give the model important background information. We explicitly instruct the model that the input text is a series of short segments separated by a special character, and that the output must follow the same structure. This line-by-line processing is critical for maintaining the temporal structure of the lecture, ensuring that each translated phrase corresponds to its original timing. We also tell the model its role is "You are a professional dubbing translator" and provide constraints such as aiming for a similar syllable count. This last instruction is very important because a translation with a similar length to the original text makes the final lip-sync and voice dubbing stages work much better.

Second, we use a Chain-of-Thought (CoT) approach \cite{ wei2022chain}, asking the model to think through several steps before giving the final translation. We instruct it to first analyze the text's tone, rhythm, and natural spoken equivalents. This step-by-step thinking process encourages the model to produce a more thoughtful and context-aware translation, rather than a simple word-for-word conversion. For instance, when translating a culturally specific idiom, the CoT process encourages the model to find a functional equivalent in the target language, rather than performing a literal, and often nonsensical, translation.

By combining these strategies and processing the text segment by segment, our translation module produces a script that is not only linguistically accurate but is also temporally aligned and primed for the subsequent stages of voice synthesis and lip synchronization, forming a critical link in our end-to-end pipeline.

\subsubsection{Creating the Voice with Accent Preservation}

Once we have a high-quality translated transcript, the next challenge is to turn this text back into speech. However, a standard, robotic text-to-speech voice would break the authentic connection with the instructor and make the lecture feel impersonal. Our goal is to create a translated voice that preserves the speaker's original accent and identity, keeping the lecture comfortable and professional.

To solve this, our system uses a technique called voice cloning. We employ a modern Text-to-Speech (TTS) model that can learn and replicate a person's voice from a short audio sample. The model works by taking two main inputs: the translated transcript from the previous step, which tells the model what to say, and the clean original voice track obtained in \ref{sec:sub:analysis}, which is used as a reference sample to tell the model how to say it by copying the instructor's unique style.

By analyzing the reference sample, the TTS model learns the instructor's unique vocal characteristics, including their pitch, tone, and importantly, their native accent. It then generates the new audio in the target language but spoken in the instructor's own voice. The result is a translated voice track that feels authentic, as if the original instructor is the one speaking the new language.

Furthermore, we recognize that some lectures feature more than one speaker. To handle this, our system provides an option to detect and process multiple speakers. When this option is selected, the audio analysis stage identifies each unique speaker and their corresponding speech segments. The voice cloning process is then applied to each speaker individually, ensuring that the final video correctly reconstructs the conversation with each person's distinct voice.

The result of this stage is a complete, translated audio track that preserves the speaker's identity, providing the essential foundation for the final lip synchronization step.

\subsubsection{Lip Synchronization}
The final step in our pipeline is to ensure that the instructor's lip movements match the newly translated audio. This is a critical step for creating a believable and seamless final video. Without proper lip synchronization, there can be a distracting mismatch between what the viewer sees and what they hear, which can break the sense of authenticity and disrupt the learning experience.

Before running the main lip-sync model, we prepare the video. We use a face detection tool to automatically scan the entire video and identify only the segments where the instructor is visible and facing the camera. This pre-processing step is very important because it allows us to ignore parts of the video that do not need modification, such as presentation slides or video clips. This saves a significant amount of handling time and focuses only on where modifications are needed.

Second, we generate the synchronized lip movements. Once we have the specific video segments containing the instructor's face, the main lip-sync process begins. Our system uses a diffusion-based model for this task, which takes two main inputs: the original video segments that we just identified, and the newly created translated audio track from the previous step.

The model analyzes the sounds phonemes in the new audio and modifies the speaker's mouth in the video frame by frame to match those sounds realistically. The output is a series of video clips where the instructor's lip movements are now perfectly synchronized with the translated speech. These modified clips are then seamlessly reinserted into the original video timeline, replacing the original segments, to create the final, polished result.

\subsection{User Interface}

\begin{figure*}[h]
  \centering
  \includegraphics[width=\linewidth]{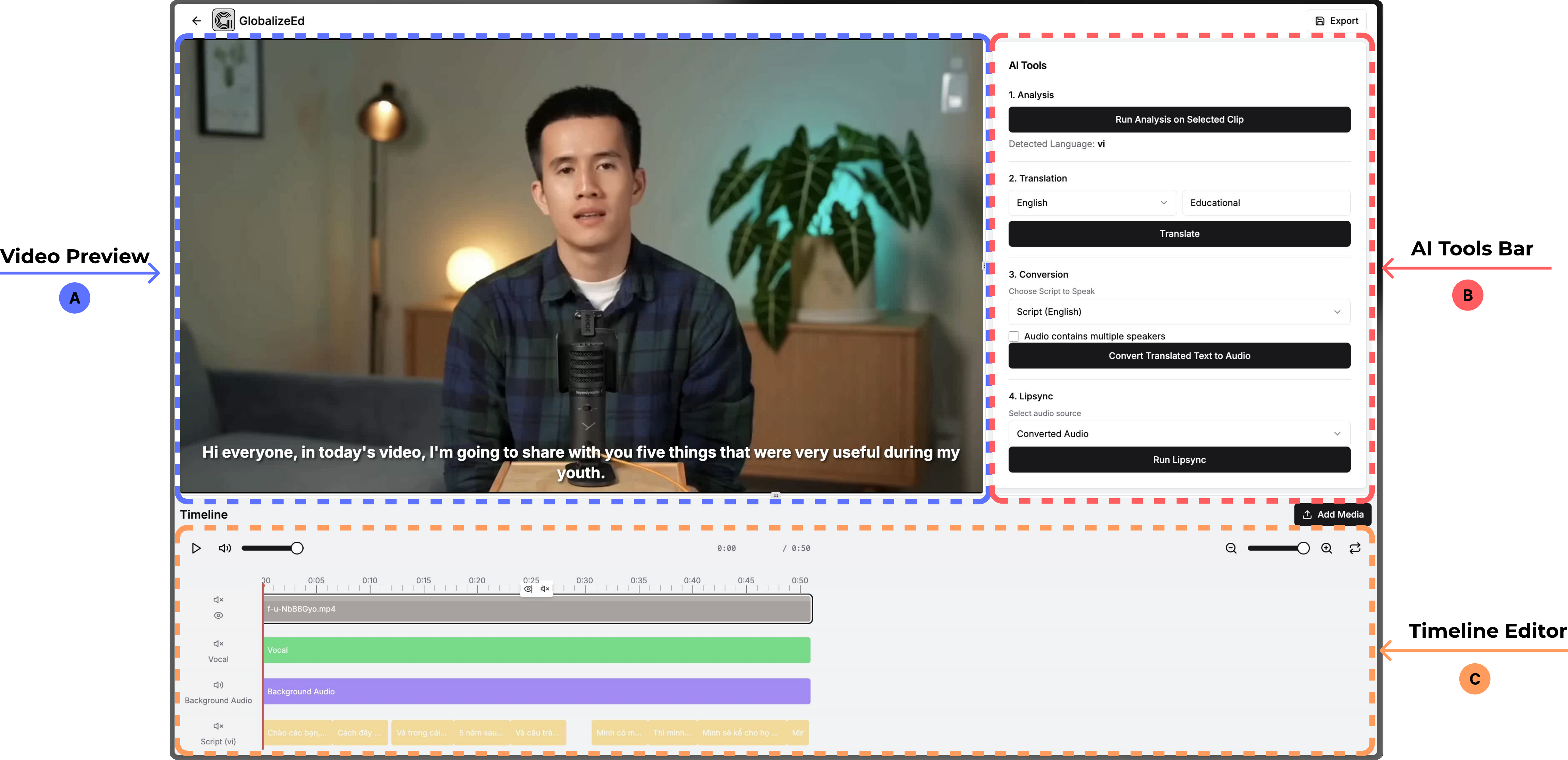}
  \caption{The user interface of the GlobalizeEd prototype. The interface is composed of three main panels: (a) a video preview window, (b) an AI Tools panel that guides the user through the step-by-step translation, and lipsync process, and (c) a timeline editor. The timeline view visualizes the separated multimodal tracks, including: video, vocal, background audio, transcript, and the translated transcript fine-grained control and visibility over the process.}
  \label{fig:ui}
\end{figure*}

The goal of GlobalizeEd is to be a professional AI tool but have an intuitive and simple interface for educators, who may not be technical experts. To achieve this, we designed an integrated user interface that organizes the entire process into a clear, manageable workflow. Figure \ref{fig:ui} shows the main interface of our prototype. The UI includes three main components, each with a specific purpose.

The AI Tools Bar Figure \ref{fig:ui}(B) is the main control center of the system. We designed it as a guided step-by-step workflow to walk the user through the process from start to finish. The process begins with Analysis, where the system automatically detects the original language of the video. Next, in the Translation step, the user can select the target language and easily modify the tone to ensure the translation style is appropriate for a lecture. The Conversion step then handles the voice cloning process. We included a useful option to handle videos with multiple speakers. Finally, the Lipsync step applies the lip synchronization model to the video. This step-by-step design prevents users from feeling overwhelmed and makes the entire process logical and easy to follow.

The Timeline Editor Figure \ref{fig:ui}(C) provides transparency and control over the process. We chose a timeline view because it is a familiar pattern for anyone who has used video editing software. After the analysis step, the timeline shows the video separated into its multimodal tracks: the original video, the isolated vocal track, the background audio track, and the generated transcript track. This visualization is important because it shows the user exactly what the AI is doing. It also gives them a clear overview of all the components of their project and provides a foundation for future features, such as editing the script directly on the timeline. On each track, the user can enable or disable any track to get a comparable perspective between original and generated content. The user can also click on each track and download it independently for any purpose.

The Video Preview window Figure \ref{fig:ui}(A) is the main canvas where users can play and review their video at any stage. It works together with the other two panels to create a seamless workflow. A typical user would upload their video, follow the steps in the AI Tools Bar, see the results appear as new tracks in the Timeline Editor, and then watch the final result in the Video Preview window before exporting the project.

\subsection{Technical Implementation}

\subsubsection{Application Architecture}

\begin{figure}[h]
  \centering
  \includegraphics[width=\columnwidth]{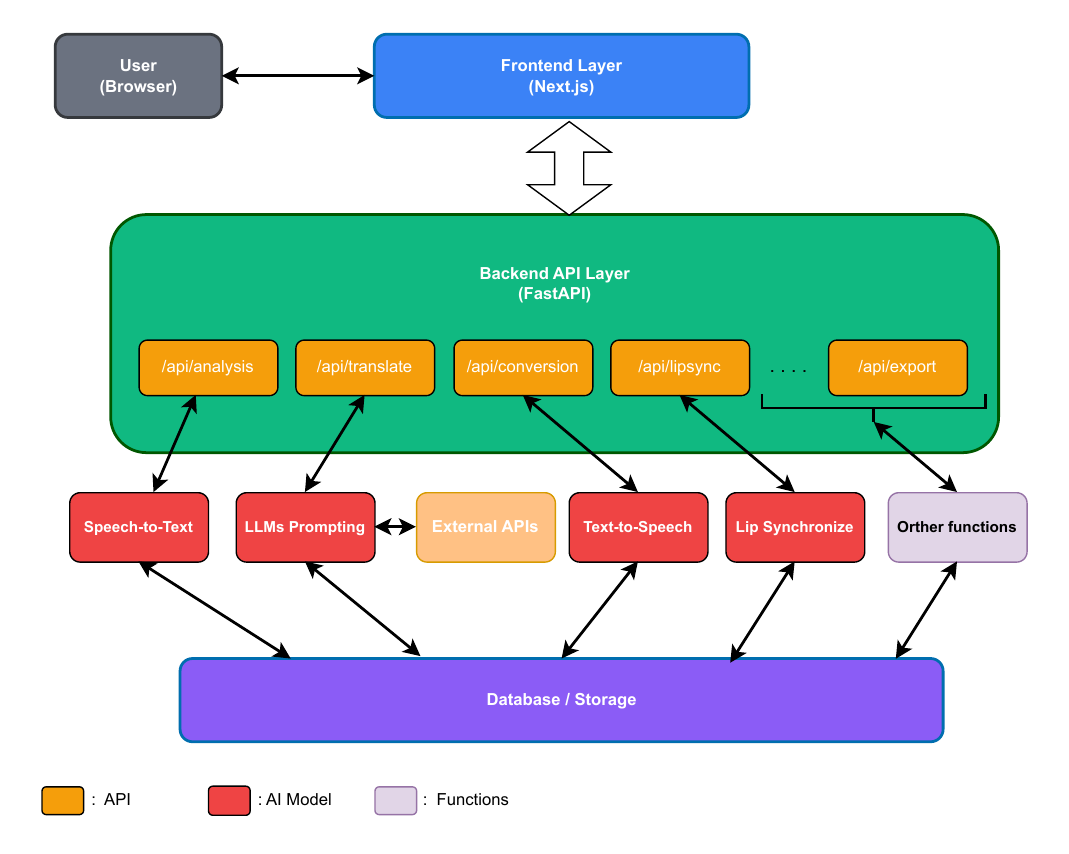}
  \caption{The technical architecture of GlobalizeEd. The user interacts with the Frontend Layer, which communicates with the Backend API Layer through REST APIs. The backend acts as an orchestrator, calling internal AI services as needed. All components interact with a central Database/Storage layer for data persistence and asset management.}
  \Description{The technical architecture of GlobalizeEd. The user interacts with the Frontend Layer, which communicates with the Backend API Layer through REST APIs. The backend acts as an orchestrator, calling internal AI services as needed. All components interact with a central Database/Storage layer for data persistence and asset management.}
  \label{fig:app_system}
\end{figure}

Our system is implemented as a Modular Monolith, a strategic architectural choice designed to balance development speed with the demands of an AI-intensive application. As shown in Figure \ref{fig:app_system}, the system is organized into distinct layers and modules, even while being a single deployable unit.

This architecture consists of a Frontend Layer that communicates with a Backend API Layer. The backend acts as an orchestrator, managing a series of internal, loosely coupled AI Modules. 
This architectural choice was strategic, offering several key advantages for a system like GlobalizeEd, which operates on a pipeline of computationally intensive AI models.

First, it enables efficient resource sharing. A major challenge in multi-model AI systems is GPU memory management. In our modular monolithic design, multiple models in the pipeline can share the same GPU memory. This approach is far more resource-efficient compared to a microservices setup, where each service would require its own dedicated hardware. Second, the architecture allows for simplified deployment and development. The entire system, along with all AI dependencies, can be packaged into a single container. This greatly simplifies deployment, reduces operational overhead, and streamlines development workflows. In practice, the unified environment also accelerates prototyping and debugging, since all components are co-located and can be tested together. Third, the design ensures low-latency communication. Modules interact through direct in-process function calls rather than over network-based APIs. Given that each model’s output directly feeds the next in the pipeline, this design eliminates network overhead and minimizes inter-module delays-resulting in faster end-to-end processing from video input to final output. Finally, the architecture establishes a foundation for future scalability. Although GlobalizeEd currently adopts a monolithic structure, its modular boundaries and well-defined interfaces make it straightforward to migrate specific components into standalone microservices if needed. As usage grows, performance-critical or frequently updated modules (such as the translation component) can be independently deployed without requiring a full architectural rewrite.

\subsubsection{Detail Implementation}

To build the GlobalizeEd prototype, we chose a Modular Monolith architecture. This approach allowed us to develop the system as a single application, which simplifies deployment and allows for efficient sharing of hardware resources. At the same time, we designed each core AI function as a separate module to ensure the system is maintainable and scalable. The technical stack and specific models used are detailed below. 

\textbf{Application Stack:} The system is implemented with a modern web stack. The Frontend is a web application built with Next.js \footnote{https://nextjs.org/}, which runs in the user's browser and communicates with the backend via REST APIs. The Backend is built with FastAPI \footnote{https://fastapi.tiangolo.com/}, a high-performance Python framework that serves as the central orchestrator for the entire AI pipeline. All data and generated assets, such as video files and transcripts, are managed by a central Storage layer. Figure \ref{fig:app_system} illustrates this layered architecture.

\textbf{AI Models:}
The AI capabilities of our system are powered by a selection of specialized models, integrated in a sequential pipeline. The process begins with Audio Separation and Transcription, where we first use MDX-Net models to isolate the instructor's vocal track. This clean audio is then fed into Faster Whisper \footnote{https://pypi.org/project/faster-whisper}, an optimized version of OpenAI's Whisper model, to generate an accurate transcript with word-level timestamps. For the Translation stage, we have been trying many LLM models, but lastly we decided to use gemini-2.5-pro from Google \cite{comanici2025gemini}

Next, for Voice Synthesis, we use Fish-Speech \cite{fish-speech-v1.4}, a high-fidelity, zero-shot Text-to-Speech (TTS) model. It takes the translated text and a short sample of the original clean voice to perform accent-preserving voice cloning. The final step is Lip Synchronization, which is handled by LatentSync \cite{li2024latentsync}, a diffusion-based model. To optimize this process, we first use the MediaPipe \footnote{https://ai.google.dev/edge/mediapipe} library to identify relevant video segments, ensuring that the computationally intensive lip-sync model is only applied where needed.

\textbf{Hardware:}
The GlobalizeEd prototype was deployed on a high-performance workstation equipped with an NVIDIA RTX 4090 GPU (24 GB VRAM), a multi-core CPU, and 32 GB of RAM. Since the system’s performance is primarily GPU-bound, we applied several optimizations to accelerate the AI pipeline. Specifically, we used FP16 precision during inference and enabled performance-oriented features.
\section{User study}
\label{user study}

\begin{table}[h]
\centering
\caption{Participant Demographics}
\label{tab:demographics}
    \begin{tabular}{llrr}
        \toprule
        \textbf{Characteristic} & \textbf{Category} & \textbf{N} & \textbf{\%} \\
        
        \toprule
        
        \multicolumn{4}{l}{\textbf{Teachers (N=18)}} \\
        \hline
        Teaching experience & 0--2 years & 4 & 22.2\% \\
         & 3--5 years & 3 & 16.7\% \\
         & 6--10 years & 6 & 33.3\% \\
         & 10+ years & 5 & 27.8\% \\
        \hline
        Educational environment & Elementary & 4 & 22.2\% \\
         & Middle School & 3 & 16.7\% \\
         & High School & 1 & 5.6\% \\
         & University & 10 & 55.6\% \\
        \hline
        Video use in teaching & A few times/month & 2 & 11.1\% \\
         & Once a week & 2 & 11.1\% \\
         & Several times/week & 11 & 61.1\% \\
         & Daily & 3 & 16.7\% \\
         
        \bottomrule
        
        \multicolumn{4}{l}{\textbf{Students (N=18)}} \\
        \hline
        Native language & Vietnamese & 9 & 50.0\% \\
         & International & 9 & 50.0\% \\
        \hline
        English proficiency & Beginner & 5 & 27.8\% \\
         & Intermediate & 10 & 55.6\% \\
         & Advanced & 3 & 16.7\% \\
        \bottomrule
    \end{tabular}
\end{table}

To evaluate the efficacy and user experience of GlobalizeEd, we conducted a mixed-methods study involving two key user groups: (1) multilingual instructors, who serve as both experts for system evaluation and potential content creators; and (2) students, who are the end-learners. Study 1 focused on understanding the needs, priorities, and evaluations of instructors regarding GlobalizeEd. Study 2 was designed to measure and compare the learning effectiveness and user experience of students when using video formats generated by our system.

These studies were designed to answer the following Research Questions:
\begin{itemize}
    \item \textbf{RQ1}: What factors do instructors prioritize most in a multimodal lecture translation tool (e.g., accuracy, vocal authenticity, or lip synchronization)?

    \item \textbf{RQ2}: How does the GlobalizeEd system impact the student learning experience (in terms of learning efficacy, cognitive load, and perceived quality) in comparison to traditional translation methods (e.g., subtitles)?

    \item \textbf{RQ3}: How do the priorities identified by instructors (Study 1) correlate with the actual benefits perceived by students (Study 2)?
    
\end{itemize}

\subsection{Study 1: Evaluation of experts}

\subsubsection{Participants}
We recruited 18 instructors (N=18) from a variety of fields and experience levels. Participants included instructors in elementary, middle, high school, and university, with teaching experience ranging from 1-2 years to over 10 years. Table \ref{tab:demographics} shows that the university level has the highest proportion at 55.6\%. In addition, the distribution of teaching experience is relatively balanced across participants with the standard deviation of \(\sim 1.29\), and most instructors (61.1\%) incorporate videos into their teaching several times per week.

\subsubsection{Procedure}

Each session with an instructor consisted of three stages:
\begin{enumerate}
    \item \textbf{Pre-task Survey}: Before interacting with the system, participants were asked to rate the importance of five key features on a 5-point scale (1 = Not important, 5 = Very important). The features included: (1) Translation Accuracy, (2) Voice Cloning, (3) Time Savings, (4) Cultural Adaptation, and (5) Lip-sync. The results of this survey are visualized in Figure \ref{fig:execptation}.

    \item \textbf{Hands-on Task}: Each instructor was given a demo and guided to use GlobalizeEd to translate a short (3-5 minute) lecture video of their own from their native language to English.

    \item \textbf{Post-task Survey}: Following the completion of the final output, we conducted a minor interview. Participants were asked to rate the resulting video on a 5-point Likert scale across six criteria: (1) Translation Quality, (2) Potential to Enhance Teaching Effectiveness, (3) Student Comprehensibility, (4) Ease of Use, (5) Vocal Naturalness, and (6) Overall Satisfaction. These ratings are shown in Figure \ref{fig:teacher_results}. The interview also aimed to gather in-depth qualitative feedback on the reasoning behind their ratings and suggestions for improvement.
    
\end{enumerate}

\subsection{Study 2: Comparative Experiment on the Student Learning Experience}

This study was designed to quantitatively compare the impact of different translation formats on the student learning experience.

\subsubsection{Participants}
We recruited 18 university students (N=18). To ensure diversity, the participant group consisted of 9 Vietnamese students and 9 international students. The English proficiency of participants was also varied, ranging from intermediate to advanced, as determined by an initial self-assessment shown in Table \ref{tab:demographics}.

\subsection{Experimental Design}

We employed a within-subjects design where each student experienced all four experimental conditions. The presentation order of the conditions was randomized.

The conditions included:
\begin{itemize}
    \item (A) Baseline: The original video with automatically generated English subtitles from YouTube.

    \item (B) Our Subtitles: The original video with English subtitles translated by our LLM.

    \item (C) Dubbing: The video dubbed using voice cloning into English without lip synchronization.

    \item (D) Full System: The video dubbed into Vietnamese with full lip synchronization.
    
\end{itemize}

\subsubsection{Materials and Measures}
 We prepared few videos with a duration of 1-2 minutes, lecture videos on neutral academic topics. Two types of videos were used: The video featured a talking-head instructor, while the other was animated. For the animated video, condition (D), Full System was omitted as there were no human lips to synchronize.
 After viewing each video, students completed a questionnaire using a 5-point Likert scale to measure the following aspects:
\begin{itemize}
    \item Learning Efficacy: The extent to which the video helped them understand and grasp the content.

    \item Cognitive Load: To assess the mental effort involved, we asked questions regarding the required level of concentration and the ease of following the lecture, based on principles from cognitive load theory \cite{van2010cognitive}.

    \item Perceived Quality: Ratings of accuracy, naturalness, and the overall viewing experience.

    \item Preference Question: After completing all conditions, students were asked: "Of the formats you just watched, which one would you prefer to learn with and why?"
    
\end{itemize}

\subsubsection{Procedure}

Each session lasted approximately 5-10 minutes. Students watched the videos under the various conditions in a randomized order. After each video, they had 2-3 minutes to answer the corresponding questionnaire. At the end of the session, they answered the final open-ended preference question.

\section{Evaluation and Results}
\label{sec:results}

In this section, we will analyze the data collected from the User Studies and present the results from our mixed-methods evaluation, which includes two separate studies. First, we present the analysis from the data collected from instructors, focusing on their vision and desires for a tool like this, as well as their initial evaluation of the application. After that, we report  the results from the experiment with students, which mainly focuses on learning effectiveness and their experience with the outputs. Finally, we examine the connection between the two user study groups.

\subsection{Results Analysis from the Experts}

The study with multilingual instructors (N = 18) aimed to quantify which features they prioritize in an educational translation tool and how they evaluate the GlobalizeEd prototype using a 5-point Likert scale.

\subsubsection{What Experts Consider Most Important}
\begin{figure}[h]
  \centering
  \includegraphics[width=\columnwidth]{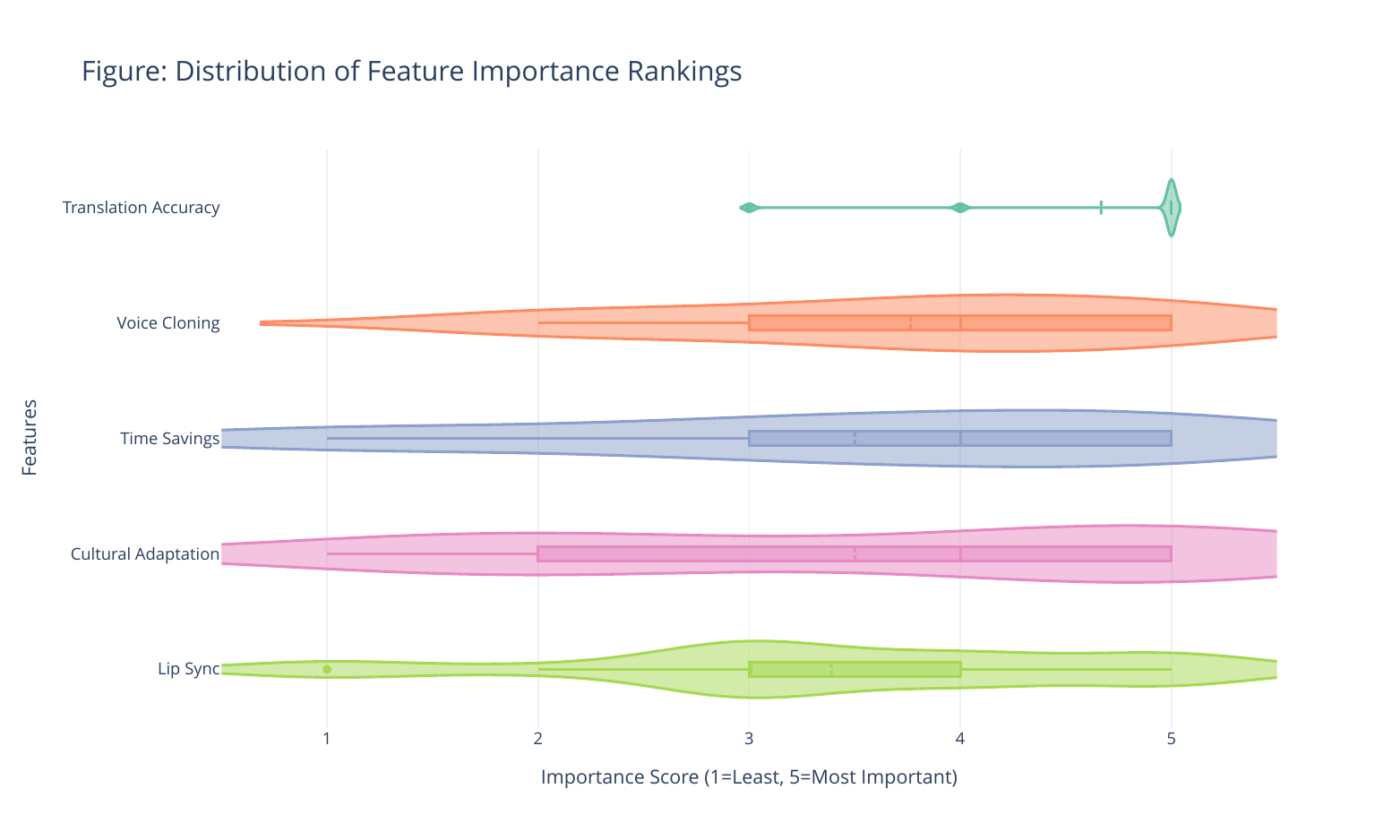}
  \caption{Feature Importance Rankings. The violin plots show the distribution of importance scores for five system features, as ranked by the instructors (1=Least Important, 5=Most Important).}
  \Description{Feature Importance Rankings. The violin plots show the distribution of importance scores for five system features, as ranked by the instructors (1=Least Important, 5=Most Important).}
  \label{fig:execptation}
\end{figure}

Before evaluating the system, we asked the instructors to rank the importance of five key features. Figure~\ref{fig:execptation} shows the distribution of these rankings.

The results revealed a clear priority order: Translation Accuracy received the highest mean score (\(\mu\)) and standard deviation (\(\sigma\)) is small \( (\mu = 4.67, \sigma = 0.686)\), showing minimal variance across participants (Figure~\ref{fig:execptation}). Following that, features that help create an authentic experience, such as Voice Cloning (\(\mu =  3.76 , \sigma=1.09\)) and Time Savings (\(\mu =  3.5 , \sigma=1.58\)), were rated as very important. This highlights that for educators, conveying the content accurately is the top priority, but maintaining the unique character and authenticity of the lecture and saving time for repair has great value.

\subsubsection{How the System Met Expectations}
\begin{figure}[h]
  \centering
  \includegraphics[width=\columnwidth]{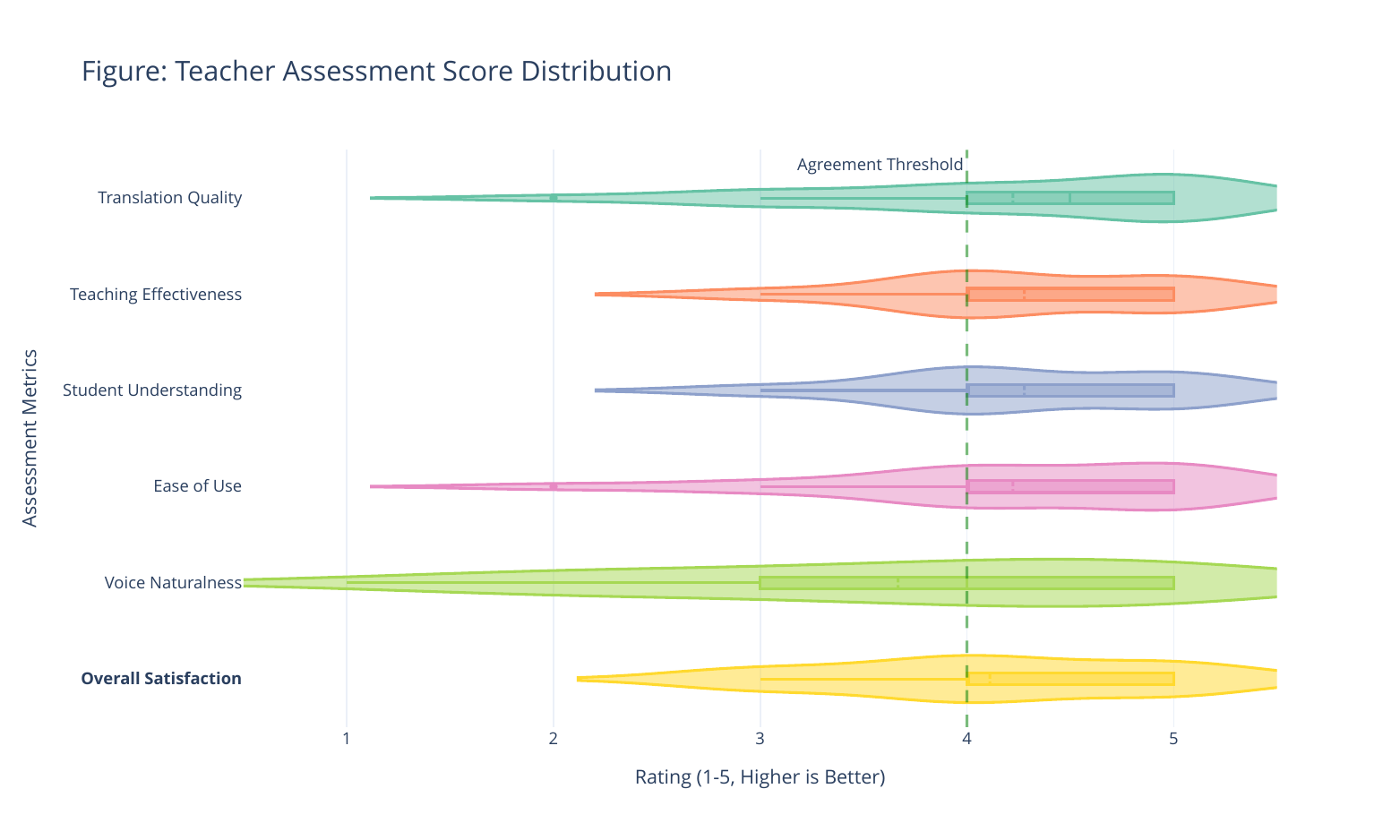}
  \caption{Teacher Assessment Score Distribution. The violin plots show the distribution of instructor ratings (1-5) for the system across six metrics. The dashed line represents the 4.0 agreement threshold.}
  \Description{Teacher Assessment Score Distribution. The violin plots show the distribution of instructor ratings (1-5) for the system across six metrics. The dashed line represents the 4.0 agreement threshold.}
  \label{fig:teacher_results}
\end{figure}

After trying the system, the instructors gave their evaluation on six key metrics. Figure~\ref{fig:teacher_results} shows these results.

The evaluation results from teachers were generally positive, with five out of six metrics scoring above 4.0, indicating strong agreement. Teaching Effectiveness and Student Understanding both received the highest mean ratings of  $M = 4.28$ ($\sigma = 0.67$), with 88.9\% of teachers rating 4 or above. Ease of Use scored $M = 4.22$ ($\sigma = 0.88$, 83.3\% agreement),  demonstrating good system usability. Translation Quality also scored  $M = 4.22$ ($\sigma = 0.94$, 77.8\% agreement) and achieved the highest median score of 4.5, indicating consistent high ratings from experienced educators.  
However, Voice Naturalness received a lower mean score of $M = 3.67$  ($\sigma = 1.28$), falling below the agreement threshold of 4.0, with only 61.1\%  of teachers rating 4 or above. This suggests that while the voice cloning feature was valued by teachers (as shown \ref{fig:execptation}), its implementation quality requires improvement. Despite this limitation, Overall Satisfaction scored $M = 4.11$ ($\sigma = 0.76$), with 77.8\% of teachers rating 4 or above,  confirming generally positive reception of the system from an expert pedagogical perspective.

\subsection{Student Evaluation}

The study with students was designed to quantitatively compare the different translation formats. These included (A) YouTube's Baseline (the original video with YouTube's automatic subtitles), (B) Our Subtitles (the original video with our LLM-translated subtitles), (C) Dubbing (our dubbed audio without lip-sync), and (D) The Full System (dubbed audio with lip-sync).

\begin{figure*}[h]
  \centering
  \includegraphics[width=\linewidth]{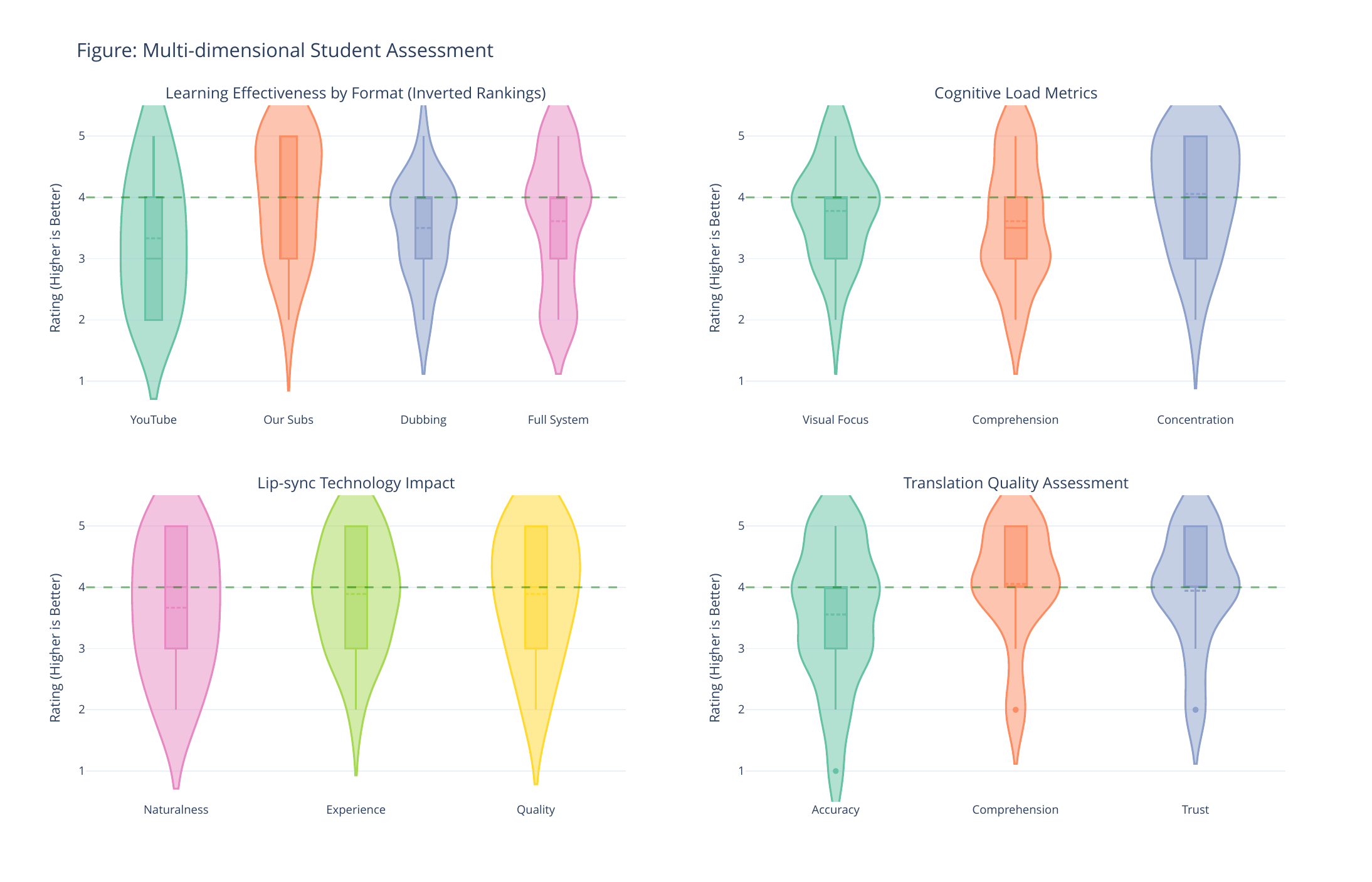}
  \caption{Multi-dimensional Student Assessment. The violin plots show the distribution of student ratings (1-5, higher is better) for four key areas: (a) Learning Effectiveness by Format, (b) Cognitive Load Metrics, (c) Lip-sync Technology Impact, and (d) Translation Quality Assessment.}
  \Description{Multi-dimensional Student Assessment. The violin plots show the distribution of student ratings (1-5, higher is better) for four key areas: (a) Learning Effectiveness by Format, (b) Cognitive Load Metrics, (c) Lip-sync Technology Impact, and (d) Translation Quality Assessment.}
  \label{fig:student_results}
\end{figure*}

We asked students to give ratings on a 5-point scale for many different aspects. Figure~\ref{fig:student_results} summarizes these multi-dimensional assessments.

The results in Figure~\ref{fig:student_results} a show interesting patterns in Learning Effectiveness. Both dubbing conditions, Dubbing (C, $Mdn = 4.0$) and Full System (D, $Mdn =4.0$), achieved median ratings of 4.0. Notably, Our Subtitles (B) also achieved $Mdn = 4.0$ and had the highest mean rating ($M = 4.00$, $\sigma = 0.97$), suggesting that our LLM-enhanced subtitles performed as well as dubbing formats. Only the baseline YouTube subtitles (A) scored lower ($Mdn = 3.0$).

The Cognitive Load Metrics (Figure~\ref{fig:student_results}b) provide insight into the dubbing experience. Concentration received the highest rating ($M = 4.06$, $Mdn = 4.0$, 72.2\% agreement), suggesting that dubbed formats allowed students to focus better on lecture content. However, Comprehension scored lower ($M = 3.61$, $Mdn = 3.5$, 50\% agreement), indicating room for improvement.

Regarding quality assessment, the Translation Quality metrics  (Figure~\ref{fig:student_results}d) showed mixed results. Translation Comprehension scored well ($M = 4.06$, 83.3\% agreement), but Translation Accuracy was lower ($M = 3.56$, 55.6\% agreement), suggesting the LLM-generated translations need refinement. The Lip-sync Technology Impact (Figure~\ref{fig:student_results}c) received moderate ratings across all metrics  ($Mdn = 4.0$ for all three aspects: Naturalness, Experience, and Quality), with approximately 60-67\% of participants rating 4 or above. While positive, these ratings indicate the lip-sync feature met but did not exceed expectations.

\subsection{Cross-Study Analysis}

\begin{figure*}[h]
  \centering
  \includegraphics[width=\linewidth]{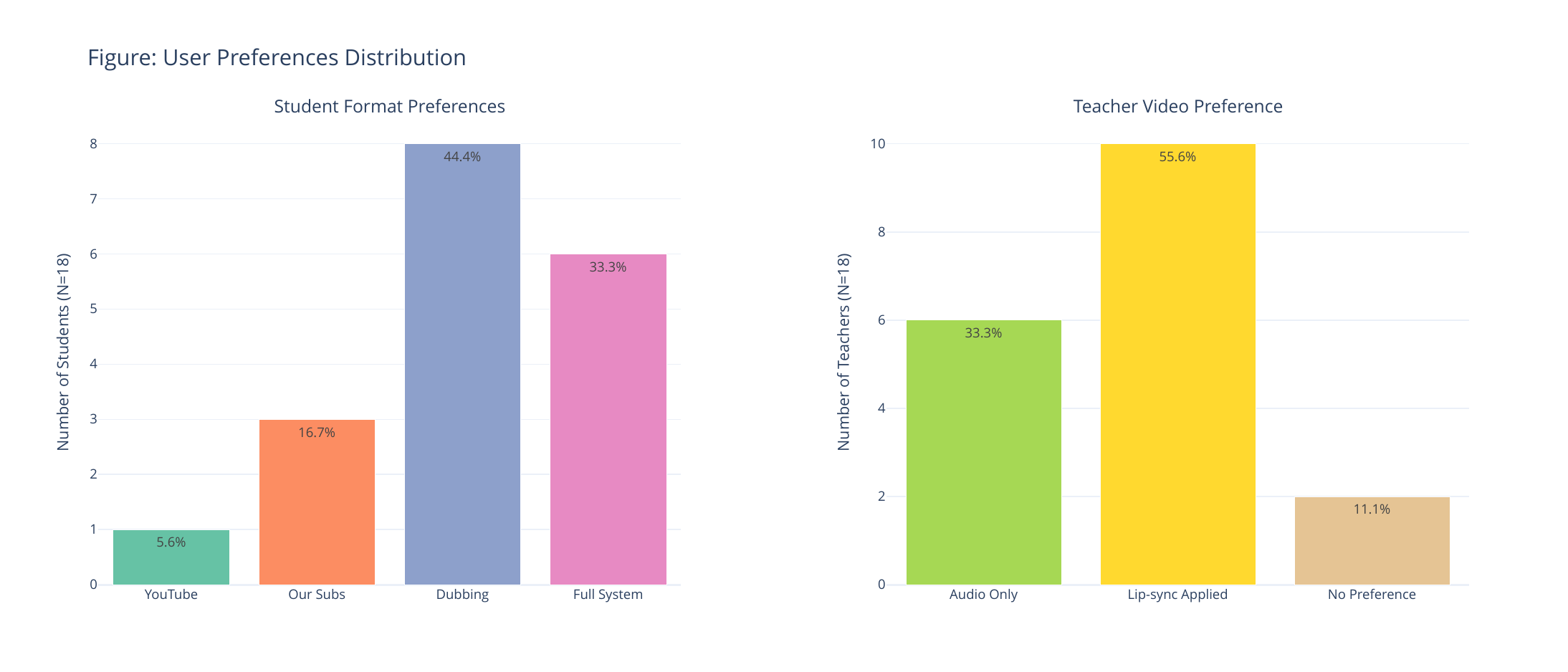}
  \caption{User Preferences Distribution. (a) Student Format Preferences: The final format preference of students after experiencing all four conditions. (b) Teacher Video Preference: The preference of instructors regarding the inclusion of the lip-sync feature.}
  \Description{User Preferences Distribution. (a) The final format preference of students after experiencing all four conditions. (b) The preference of instructors regarding the inclusion of the lip-sync feature.}
  \label{fig:cross_stu}
\end{figure*}

One of the strongest findings of this work comes from comparing the results between the two user groups.

Correlation between Teacher Priorities and Student Benefits:
\begin{itemize}

    \item Translation Accuracy as Foundation: Instructors identified      Translation Accuracy as the number one priority (Figure~\ref{fig:execptation}). This is directly reflected in the student results: they rated Translation Comprehension highly ($M = 4.06$, 83.3\% agreement), and ultimately, the vast majority (94.4\%) chose one of GlobalizeEd's formats (B, C, or D) over the      YouTube baseline (Figure~\ref{fig:cross_stu}a). This demonstrates that a high-quality translation foundation is essential for end-user adoption.
    
    \item The Value of Authenticity: Instructors ranked Voice Cloning as the second most important feature. On the student side, the results show that the dubbed formats (C and D, which use voice cloning) were most preferred, with 77.7\% of students choosing these formats. Students perceived these formats as having higher Learning Effectiveness (both achieving $Mdn = 4.0$) and reported better Concentration ($M = 4.06$). This shows that authenticity through voice cloning has a direct and measurable impact on the learning experience.

    \item Different Perspectives on Lip-sync: Instructors ranked Lip-sync with the lowest importance when asked abstractly (Figure~\ref{fig:execptation}), yet when choosing the final product, 55.6\% preferred to include it versus 33.3\% who preferred audio-only      (Figure~\ref{fig:cross_stu}b). Among students, preferences were more evenly split: 44.4\% preferred Format C (dubbing without lip-sync) versus 33.3\% who preferred Format D (with lip-sync). While lip-sync received moderate ratings ($Mdn = 4.0$ across all metrics), it did not show a statistically significant improvement in perceived quality ($p = 0.73$). This suggests that lip-sync, while appreciated by some users for its professional appearance, is not a decisive factor for all learners. The qualitative feedback revealed that when lip-sync is well-executed, it enhances naturalness and reduces distraction. However, poor synchronization can have the opposite effect, making it more important to prioritize quality over mere presence of this feature.
    
\end{itemize}

\subsection{Qualitative Insights}
To understand the reasons behind our quantitative results, we analyzed the open-ended comments from students and instructors \cite{braun2006using}. This analysis highlighted three key themes that explain how users experienced and thought about the different formats.

\subsubsection{Two Learning Preferences: Active Learning with Subtitles vs Easy Viewing with Dubbing}
Student feedback showed a clear split in preferences, pointing to two different ways of learning.

One group of participants preferred the subtitled formats (A and B). They saw subtitles as a tool for \textbf{active learning}. These students used the text to learn English, understand difficult terms, and quickly scan the content.
\begin{quote}
\textit{I preferred Format B because the improved subtitles were clearer and more accurate. They helped me understand difficult vocabulary and academic terms without losing focus on the visuals.}
--- [P013]
\end{quote}
\begin{quote}
\textit{Because I think a video with subtitles helps me learn English easier.}
--- [P012]
\end{quote}

In contrast, the larger group preferred the dubbed formats (C and D), seeking an \textbf{effortless viewing experience}. For them, reading subtitles required extra effort and took their focus away from the lecture. Keeping the instructor's original voice was important because it felt more familiar and engaging.
\begin{quote}
\textit{Easier to understand, but still retains the speaker's emotions.}
--- [P010]
\end{quote}
\begin{quote}
\textit{It looks more natural and the information is easier to assimilate.}
--- [P018]
\end{quote}
This theme clearly explains the quantitative data. While good subtitles are helpful, dubbed formats were more effective at lowering the mental effort required, allowing students to concentrate fully on the lecture.

\subsubsection{The Critical Role of Synchronization: From Distraction to Professionalism}
The qualitative feedback strongly emphasized the critical role of audio-visual synchronization, especially lip movements. When the audio and video were out of sync, it was a major distraction that broke the viewing experience. Both students and instructors described this mismatch as "annoying" and "strange".
\begin{quote}
\textit{Format C was more distracting than Format B as the face and dubs do not match. It was better when dubs and lips were aligned.}
--- [P017]
\end{quote}
\begin{quote}
\textit{(Instructor) Make sure facial expressions match the voice—I noticed some repetitive movements with no sounds coming together, which was very annoying and strange.}
--- [T018]
\end{quote}

On the other hand, when the synchronization was done well, it made the video feel much more natural and professional. This improved not only how the video looked but also made it feel more trustworthy, helping viewers stay focused.
\begin{quote}
\textit{It feels natural and not forced, better than usual!}
--- [P009]
\end{quote}
This theme directly supports our cross-study findings. It shows why students rated the ``Full System'' so highly and why instructors realized the importance of lip-sync after seeing the final result. Synchronization is not just a visual polish; it is essential for creating a professional and believable final product.

\subsubsection{Instructor Feedback: The Need for Clearer Controls and Workflow Integration}
Instructors' comments focused on the practical use of GlobalizeEd as a daily tool. Their feedback highlighted the importance of a user-friendly interface and how the tool fits into their teaching process.

A key point of friction was the discoverability of features. For example, several instructors requested the ability to download or export the materials they created.
\begin{quote}
\textit{I wish this app could have the download function that allows me to at least download the translated script, and if I can download the lipsynced video, that will be the best.}
--- [T004]
\end{quote}
\begin{quote}
\textit{Also, I want to export the scripts in both languages into a .doc file.}
--- [T009]
\end{quote}
Interestingly, the prototype does allow for the download of individual tracks from the timeline editor. This feedback, therefore, highlights not a missing feature, but a crucial \textbf{usability issue}: the function was not intuitive or visible enough for users to find. It underscores the need for a clearer user interface where essential actions like exporting are readily accessible.

Furthermore, instructors provided other valuable suggestions, such as improving processing speed [T008] and refining the cloned voice to better match the pronunciation of the target language [T010]. This feedback shows that for a tool like GlobalizeEd to be widely adopted, it must not only be powerful but also easy to use and flexible enough to fit into the real-world workflows of educators.
\section{Discussion and Conclusion}
\label{sec:disc}

In this research, the main goal was to study the needs and reactions of teachers and students when applying AI to translate academic videos. Through this application, lectures in niche fields can be shared more widely, which is why we named the application GlobalizeEd. By conducting a mixed-methods experiment with both instructors and students, we found that our approach significantly improved learning comprehension and reduced cognitive load compared to traditional subtitles. The results also highlight the importance of balancing technological innovation with user control and authenticity. In this final section, we will discuss the broader implications of these findings, address important ethical considerations, point out the limitations of the research, and conclude with future work directions.

\subsection{Lessons for Design and Learning}

The results we collected have led to a few important design lessons for future educational tools.

\textbf{First, Accuracy is the foundation, but Authenticity makes the difference.} Instructors considered accurate translation to be the most critical condition for an academic application. But it was the features related to authenticity, like preserving the voice, that helped create a less strenuous and more immersive learning experience. This means a good tool needs to be accurate first, and then features that preserve authenticity can reduce cognitive barriers and improve engagement, making the content more accessible. This insight aligns with research showing that preserving the instructor's voice enhances their social presence \cite{garrison1999critical}, fostering a stronger connection that contributes to the learning experience.

\textbf{Second, Human-in-the-loop.} The feedback of the instructors was very clear: for important academic content, they must have the ability to review and edit the AI's results. This shows that AI should be a powerful assistant that supports the work process, which was also the direction we aimed for. The instructor must be the user who can then provide an expert perspective to judge if the result meets the standard for use.

\textbf{Finally, Synchronization is the deciding factor for perceived quality.}
Although instructors did not rank lip-sync high initially, both they and the students felt very uncomfortable when the audio and video did not match. This shows that synchronization is a basic factor; when it is good, no one notices, but when it is bad, it ruins the whole experience and causes distraction from learning.

\subsection{Ethical and Responsibility Issues}

The technologies in GlobalizeEd, like voice cloning and lip synchronization, are powerful, and we need to think carefully about ethical issues.

\textbf{Deepfake, Voice Ownership, and Consent.} The face and voice are a part of personal identity. Copying someone's voice needs their clear permission. Our system is designed to be used by the instructors themselves on their own content, which is a correct step. However, there need to be clear rules about who is allowed to use another person's voice. This also emphasizes once again the need for a Human-in-the-loop approach in academic products like this to ensure privacy and ethics. At the same time, bad actors could do similar things, a recognized challenge for privacy and society \cite{chesney2019deep}. It leads us to consider about security measures such as digital signatures to prevent misuse.

\textbf{Bias in AI models.} AI models can learn the biases that exist in their data. For example, they might not translate local dialects well. We need to continuously check to make sure these models work fairly for everyone.

\subsection{Limitations and Future Development}

Our research still has some limitations. The number of participants was not very large, and the experiment was done with short videos. In addition, we have only tested translation from a source language to English instead of both ways, and the AI technology is sometimes still not perfect.

These limitations open up interesting directions for the future:

\begin{itemize}
    \item \textbf{Real-world testing:} Bring GlobalizeEd into a real classroom for a whole semester to see its long-term impact.
    \item \textbf{Building editing tools:} Based on instructor feedback, create tools that help them easily fix translation errors.
    \item \textbf{Real-time dubbing:} Research whether this technology can be applied to live lectures.
    \item \textbf{More research on Lip-sync:} Investigate more deeply to know when a nearly perfect lip-sync is better or worse than having none at all.
\end{itemize}

These future directions could contribute significantly to research on applying AI in professional academic environments.

\subsection{Conclusion}

In summary, this work introduces GlobalizeEd, a system that shows the potential of AI dubbing while still preserving the speaker's identity in education. By creating an authentic and multimodal learning experience, we have shown that it is possible to help learners understand better, reduce cognitive load, and create a stronger connection between instructors and students across language barriers. These results provide a solid experimental foundation and offer clear design lessons for the next generation of human-centered AI tools.

\bibliography{content/refs}
\bibliographystyle{ACM-Reference-Format}

\end{document}